\documentclass[conference]{IEEEtran}
\IEEEoverridecommandlockouts
\usepackage{cite}
\usepackage{amsmath,amssymb,amsfonts}
\usepackage{algorithmic}
\usepackage{graphicx}
\usepackage{textcomp}
\usepackage{xcolor}
\def\BibTeX{{\rm B\kern-.05em{\sc i\kern-.025em b}\kern-.08em
    T\kern-.1667em\lower.7ex\hbox{E}\kern-.125emX}}

\usepackage{url}
\usepackage{xspace}
\usepackage{tikz, pgfplots, pgfplotstable}
\usepackage{paralist}
\usepackage{subfig}
\usetikzlibrary{
  shapes.geometric,
  arrows,
  external,
  pgfplots.groupplots,
  matrix
}

\newcommand{\acro}{{\textit{UCCA}}\xspace}
\newcommand{\acrotitle}{{{ \textbf{UCCA}}}\xspace}

\newcommand{\region}{{\textit{UCC}}\xspace}
\newcommand{\regions}{{\textit{UCC}s}\xspace}
\newcommand{\regmin}{$\region_{min}$\xspace}
\newcommand{\regmax}{$\region_{max}$\xspace}
\newcommand{\CR}{\textit{CR}\xspace}
\newcommand{\BP}{\textit{BP}\xspace}
\newcommand{\RTN}{\textit{$RET_{exp}$}\xspace}
\newcommand{\daddr}{$D_{addr}$\xspace}
\newcommand{\wen}{$W_{en}$\xspace}
\newcommand{\PC}{\textit{PC}\xspace}
\newcommand{\rst}{\textit{reset}\xspace}
\newcommand{\SP}{\textit{SP}\xspace}
\newcommand{\IRQ}{\textit{$IRQ_{jmp}$}\xspace}
\newcommand{\OP}{\textit{$OP_{ret}$}\xspace}
\newcommand\adv{\ensuremath{\sf{\mathcal Adv}}\xspace}

\usepackage{enumitem}

\newlist{myitemize}{itemize}{1}
\setlist[myitemize]{
    label=$\bullet$,
    align=left,
    leftmargin=*,
    nosep,
}

\newlist{myenumerate}{enumerate}{1}
\setlist[myenumerate]{
  label=\arabic*.,
  align=left,
  leftmargin=*,
  nosep,
}

\newcounter{revs}

\begin{document}

\title{\huge \acrotitle: A Verified Architecture for Compartmentalization of Untrusted Code Sections in Resource-Constrained Devices}

\author{\IEEEauthorblockN{Liam Tyler}
\IEEEauthorblockA{
\textit{Rochester Institute of Technology}\\
lgt2621@rit.edu}
\and
\IEEEauthorblockN{Ivan De Oliveira Nunes}
\IEEEauthorblockA{
\textit{Rochester Institute of Technology}\\
ivanoliv@mail.rit.edu}
}

\maketitle

\begin{abstract}
Micro-controller units (MCUs) implement the de facto interface between the physical and digital worlds. As a consequence, they appear in a variety of sensing/actuation applications, from smart personal spaces to complex industrial control systems and safety-critical medical equipment. While many of these devices perform safety- and time-critical tasks, they often lack support for security features compatible with their importance to overall system functions. This lack of architectural support leaves them vulnerable to run-time attacks that can remotely alter their intended behavior, with potentially catastrophic consequences. In particular, we note that MCU software often includes untrusted third-party libraries (some of them closed-source) that are blindly used within MCU programs, without proper isolation from the rest of the system. In turn, a single vulnerability (or intentional backdoor) in one such third-party software can often compromise the entire MCU software state. 

In this paper, we tackle this problem by proposing, demonstrating security, and formally verifying the implementation of \acro: an \underline{U}ntrusted \underline{C}ode \underline{C}ompartment \underline{A}rchitecture. \acro provides flexible hardware-enforced isolation of untrusted code sections (e.g., third-party software modules) in resource-constrained and time-critical MCUs. To demonstrate \acro's practicality, we implement an open-source version of the design on a real resource-constrained MCU: the well-known TI MSP430. Our evaluation shows that \acro incurs little overhead and is affordable even to lowest-end MCUs, requiring significantly less overhead and assumptions than prior related work.
\end{abstract}

\begin{IEEEkeywords}
Memory Protection, Compartmentalization, Embedded Systems
\end{IEEEkeywords}

\section{Introduction}\label{sec:intro}

Embedded systems have become critical components of many applications, including cyber-physical systems (CPS) and the Internet-of-Things (IoT). Normally, these devices feature one or more resource-constrained micro-controller units (MCUs) responsible for interfacing with the physical world (i.e., sensing and actuation). MCUs are often designed to minimize cost, size, and energy consumption. As such, they usually run software in place (physically from program memory) and lack virtual memory and other forms of isolation commonly found in higher-end devices.


Due to their budgetary limitations, MCUs are often left vulnerable to run-time exploits~\cite{szekeres2013sok,roemer2012return,checkoway2010return,bletsch2011jump,shacham2007geometry}  (for instance, triggered by buffer overflow vulnerabilities~\cite{one1996smashing,smith1997stack}). Run-time attacks allow an adversary to remotely alter the intended behavior of a program during its execution. Without proper isolation, a single run-time vulnerability could give an adversary full control over the device~\cite{cve_2022, print_nightmare}. This can be used to spoof sensor data, bypass safety checks, ignore remote commands, and ignore scheduled task deadlines. For instance, a compromised patient-monitoring system implemented using MCUs could fail to alert medical personnel in case of an emergency~\cite{abera2016c} 
or cause a denial of service~\cite{antonakakis2017understanding}. Similarly, vulnerable industrial control sensors could be used to run machines at unsafe speeds and damage equipment (e.g., as in the Stuxnet attack~\cite{falliere2011w32}).

Some MCUs (e.g., in the ARM Cortex-M family~\cite{arm_cortex_m}) support rudimentary isolation to mitigate run-time attacks. Privilege levels~\cite{clements2017protecting,clements2018aces} allow the MCU to run software as either privileged or unprivileged. The MCU also restricts how privileged code can be called. This enables the isolation of privileged code from unprivileged code. Thus, unprivileged run-time vulnerabilities cannot access the privileged functionality. While useful, this mitigation is still limited, because unprivileged vulnerabilities can still compromise all unprivileged code. Similarly, privileged vulnerabilities can reach all privileged and unprivileged software. This means vulnerabilities within a privileged function still result in a full system compromise. 

Memory Protection Units (MPUs) allow for isolation between the privileged and unprivileged layers and further restrictions within each layer, by enforcing read, write, and execute permissions to a fixed number of memory segments. This allows more restricted compartments within the unprivileged layer, however, MPUs are configurable by the privileged software. As such, they cannot restrict privileged code, as any compromised privileged code could misconfigure the MPU.
To make matters worse, the privileged layer must implement several low-level system functions, including all Interrupt Service Routines (ISRs) and respective drivers~\cite{almakhdhub2020mu,clements2018aces}, Direct Memory Access (DMA) management~\cite{mera2022d}, real-time task scheduling~\cite{freeRTOS}, and more. This contributes to a large and complex Trusted Computing Base (TCB) that often relies on multiple untrusted third-party software modules and libraries. 
MPU-based protection often also requires disabling interrupts for unprivileged software creating a conflict between real-time requirements and security for the MCU software. 

Motivated by this pressing issue, we propose, design, implement, and formally verify \acro: an \underline{U}ntrusted \underline{C}ode \underline{C}ompartment \underline{A}rchitecture.
\acro is a lightweight hardware-based memory isolation method that enables the definition of arbitrary-sized memory segments for untrusted code (e.g., third-party software) at device loading time (i.e., whenever physically programmed via USB, J-TAG, etc.). At run-time, \acro monitors CPU signals to actively prevent malicious behavior within the untrusted sections from escalating to the rest of the MCU.

Unlike current bare-metal approaches (e.g., MPUs) that isolate trusted functionality from the rest of the software, \acro instead isolates untrusted code. Since run-time attacks typically originate from well-known code sections (e.g., I/O functions or third-party libraries), untrusted code sections can be identified pre-deployment. Through isolation, \acro limits the reach of exploits to their own context. Attempts to obtain similar guarantees with existing hardware lead to large memory and run-time overheads, limits its applicability to unprivileged code only, and may require disabling interrupts preventing asynchronous event handling (see Section \ref{mpucca} for details). In contrast, \acro can isolate untrusted privileged code (such as drivers) and does not require disabling interrupts to enforce isolation. \acro also enables finer-grained isolation that can be used jointly with existing hardware to further isolate unprivileged applications from their own untrusted code sections and third-party libraries. \acro is designed as a hardware monitor that runs independently and in parallel with the MCU core. Therefore, no software (including privileged code) can misconfigure \acro's protections at run-time. Furthermore, \acro incurs little execution time overhead (for marshaling data into isolated compartments) and maintains support for interrupts.
In sum, this paper's anticipated contributions are three-fold:

$\bullet$ Proposal and design of a lightweight hardware-based architecture for isolation of untrusted code sections in resource-constrained MCUs. This prevents the escalation of run-time vulnerabilities to the entire system. \acro includes support for the isolation of untrusted interrupts and untrusted privileged code sections.

$\bullet$ Implementation and formal verification of \acro atop an open-source version \cite{girard_2009} of the well-known TI MSP430 MCU. \acro's prototype is publicly available at~\cite{anonymous642}.

$\bullet$ Evaluation of \acro prototype and comparison to related approaches~\cite{koeberl2014trustlite,noorman2013sancus,almatary2022compartos} in terms of hardware overhead. Along with \acro's open-source release, we implement sample attack programs, that show how their escalation is detected and prevented by \acro in practice. 

\section{Background}

\subsection{Scope of MCUs}\label{sec:scope}

This work focuses on resource-constrained embedded MCUs. These are single-core devices, executing instructions physically from program memory (i.e., at ``bare metal``), and lacking a Memory Management Unit (MMU) to support virtual memory. We target these devices because an architecture that is simple and cost-effective enough for the lowest-cost MCUs is adaptable for higher-end devices with higher hardware budgets (whereas the reverse is often more challenging). In addition, the relative simplicity of these devices enables us to reason about them formally and verify \acro security properties. With these premises in mind, we implement our \acro prototype atop the TI MSP430; a well-known low-end MCU. 
This choice is also motivated by the availability of an open-source version MSP430 hardware from OpenCores \cite{girard_2009}. Nevertheless, \acro's design and assumptions are generic and should also apply to other MCUs.


\subsection{Linear Temporal Logic (LTL) \& Formal Verification}

Computer-aided formal verification typically involves three steps. First, the system of interest (e.g., hardware, 
software, protocol) is described using a formal model,  e.g., a Finite State Machine (FSM). 
Second, properties that the model should satisfy are formally specified. Third, the system model is checked against 
these formally specified properties. This can be done via Theorem 
Proving~\cite{loveland2016automated} or Model Checking~\cite{clarke2018model}.
We use the latter to verify \acro's implementation.

We formally specify desired \acro properties using Linear Temporal Logic (LTL) and implement \acro hardware as FSMs using the Hardware Description Language (HDL) Verilog~\cite{verilog}. Hence, \acro's hardware FSM is represented by a triple: $(\sigma, \sigma_0, T)$, 
where $\sigma$ is the finite set of states, $\sigma_0 \subseteq \sigma$ is the set of possible initial states, and 
$T \subseteq \sigma \times \sigma$ is the transition relation set, which describes the set of states that can be reached 
in a single step from each state.

To verify the implemented hardware against the LTL specifications we use the popular model checker NuSMV~\cite{cimatti2002nusmv}. For digital hardware described at Register Transfer Level (RTL) (the 
case in this work) conversion from HDL to NuSMV models is simple. Furthermore, 
it can be automated~\cite{irfan2016verilog2smv} as the standard RTL design already relies on describing 
hardware as FSMs. LTL specifications are useful for verifying sequential systems. In addition to propositional 
connectives, conjunction ($\land$), disjunction ($\lor$), negation ($\neg$), and implication ($\rightarrow$), 
LTL extends propositional logic with {\bf temporal quantifiers}, thus enabling sequential reasoning. Along with the standard future quantifiers, \acro's verification also uses Past-Time LTL~\cite{lichtenstein1985glory, cimatti2002nusmv} to reason about past system states. Specifically, \acro formal specifications and respective verification 
rely on the following LTL temporal quantifiers:
\begin{myitemize}
        \item \textbf{X}$\phi$ -- ne\underline{X}t $\phi$: holds if $\phi$ is true at the next system state.
        \item \textbf{G}$\phi$ -- \underline{G}lobally $\phi$: holds if for all future states $\phi$ is true.
        \item $\psi$\textbf{W}$\phi$ -- $\psi$ \underline{W}eak Until $\phi$: holds if $\psi$ is true for \textit{at least} all states until $\phi$ becomes true or $\psi$ is globally true if $\phi$ never becomes true. 
        \item \textbf{Y}$\phi$ -- \underline{Y}esterday $\phi$ (a.k.a. Previous $\phi$): holds if $\phi$ was true in the previous system state.
\end{myitemize}
\subsection{Run-Time Exploits \& Software Isolation}\label{sec:bg_attacks}

Run-time software attacks allow an adversary (\adv) to remotely alter the intended behavior of a program. The majority of program instructions execute sequentially, however so-called {\bf branching instructions} (e.g.: function calls, returns, if statements, and loops) can alter this sequence. Thus branching instructions define the program's intended {\bf control flow}. If certain vulnerabilities are present, \adv can hijack these instructions and change the software's intended behavior. For example, buffer overflows~\cite{one1996smashing,smith1997stack} overrun a buffer's allocated memory to corrupt adjacent stack memory and potentially the current function's return address. As such, \adv can craft malicious oversized buffer inputs, overwrite return addresses, and force a jump to some \adv-defined address. Consequently, this leads to well-known attacks such as control flow hijacking~\cite{schuster2015counterfeit,evans2015control}, code injection~\cite{younan2012runtime,francillon2008code}, and Return Oriented Programming (ROP)~\cite{roemer2012return,checkoway2010return,bletsch2011jump,shacham2007geometry}. For an overview of run-time software vulnerabilities and their consequences see~\cite{szekeres2013sok}.

The recurrence of run-time exploits has led to various mitigation (see Section~\ref{sec:rw}). Among them, isolation techniques are the predominant method to prevent programs from interfering with each other. In particular, they aim to protect a given process from tampering by another malicious/compromised task executing on the same device. Higher-end devices (e.g., general-purpose computers and servers) rely on virtual memory to enforce inter-process isolation. On these devices, unprivileged processes (typically all processes but the Operating System (OS)) can only stipulate memory accesses via virtual addressing. An MMU in the CPU translates each virtual access to a physical address in real-time. These translations are only configurable by privileged software (typically the OS). Therefore as long as the MMU is securely configured, unprivileged processes cannot interfere with each others' control flow, code, or data. Notably, MMU-based isolation assumes the OS is vulnerability-free. This implies a large TCB, often including low-level code (i.e., device drivers), and has led to numerous attacks on OS implementations\cite{eternal_blue,cve_42719,cve_38040}.

Regardless of their benefits or shortcomings, the hardware cost of virtual memory and MMU-based isolation is prohibitive for MCUs. Lower-end MCUs often have no support for isolation (e.g., TI MSP430 and AVR ATMega) whereas higher-end MCUs (e.g., some ARM Cortex-M MCUs) feature less expensive MPUs. MPUs are hardware monitors that configure physical memory regions with different read, write, and execute permissions for privileged and unprivileged software. MPUs can protect security-critical code against tampering by enforcing (i) read-only permissions for critical code sections; and (ii) data execution prevention for data segments. Similar to MMUs, MPUs are configured by privileged software (e.g., an embedded OS such as FreeRTOS~\cite{freeRTOS}). Thus, MPUs must also trust the OS, as the OS can freely configure the MPU.

Some higher-end MCUs are also equipped with TrustZone-M \cite{ARM-TrustZone}. TrustZone is an architectural extension that divides MCU hardware, software, and data into a Secure and Non-Secure world. The Secure world is an isolated execution environment for security-critical software. The Secure world can only be called from the Non-Secure world through secure entry points called Non-Secure Callables (NSCs). To enable this separation, TrustZone adds new hardware extensions to the MCU. The Secure Attribution Unit (SAU) and Implementation Defined Attribution Unit (IDAU) \cite{Armv8_M_TZ_spec} mark memory as Secure, Non-Secure, and Non-Secure Callable. This assigns the memory to the corresponding world and marks it as an NSC respectively. The IDAU defines a base memory configuration that the SAU can overwrite to elevate their definitions. While the SAU and IDAU divide memory between two worlds, they do not provide further separation within each world nor prevent vulnerabilities in the Secure world from compromising the Non-Secure world. As such, the SAU and IDAU enforce configurations defined by an additional level of privilege.

We note that the premise of existing controls is that security-critical sections can be determined \textit{a priori}. \textit{\acro (this work) is rooted in the different and complementary premise that untrusted code segments, i.e., those more likely to contain software vulnerabilities can also be enumerated a priori}. We stress that this does not require that \acro pinpoints/identifies vulnerabilities themselves (a much harder task) but rather allows the definition of ``less trusted'' code sections. As discussed earlier, run-time attacks typically originate from well-known code sections e.g., low-level I/O manipulation exposed to malformed/malicious inputs and third-party (often closed-source) code. Thus these components are good candidates for compartmentalization in \acro. Once untrusted code segments are defined, \acro prevents attacks in these regions (if any) from escalating to the rest of the system. Therefore, \acro can work in tandem with existing hardware to not only protect security-critical code from the rest of the MCU software but also ensure that likely vulnerable code segments, if/when exploited, cannot escalate to the rest of the system. Importantly, \acro's design allows isolation within privileged software for increased protection even against privileged vulnerabilities.

\section{\acrotitle Overview}
\label{sec:overview}

\acro is a hardware monitor that isolates untrusted code compartments (\regions) from the rest of the system. 
What constitutes untrusted code varies with application domains and developer-defined security policies. As such, \regions are flexible to allow for different isolation cases. \regions contain executables and are defined by their first and last addresses in physical memory; namely \regmin and \regmax (Recall from Section~\ref{sec:scope} that MCUs execute instructions in-place, physically from program memory). \region locations in memory are configurable and can have arbitrary size. All \region definitions ((\regmin,\regmax) pairs) are stored in a reserved and protected part of physical memory denoted the  ``Configuration Region'' (\CR). Their values are loaded to \CR when the MCU is physically programmed/flashed and \acro prevents \CR from being overwritten at run-time. Thus, once defined, \regions cannot be changed or disabled by any software.

To isolate each \region, \acro monitors CPU signals to enforce two properties, Return Integrity and Stack Integrity. Return integrity prevents invalid returns (as well as any other malicious jumps) from \regions. Whenever execution enters a \region, \acro saves a copy of the return address. Then, when \region finishes running, \acro enforces that execution returns to this previously saved value. This prevents any control flow attacks within \region from escalating to the rest of the system. Stack integrity creates an isolated stack frame for each \region. This isolated frame allows code within \region to write to the stack and heap while preventing modifications to stack memory belonging to functions external to \region. Stack integrity also ensures the stack pointer is properly set when returning from \region. This stops attempts to corrupt data in use by other functions in the same device.

Despite these restrictions, \regions remain interruptable. If a \region is interrupted, \acro loosens return integrity to allow execution to jump to the associated ISR. Once outside the \region, stack integrity is disabled allowing the interrupt to edit the stack as needed. While the interrupted, \acro maintains the saved return address and isolated stack frame. Then when execution returns to \region, return and stack integrity are re-enforced. A malicious interrupt could abuse this behavior to break \acro's protections. Nonetheless, \acro allows any untrusted ISR to also be confined within a dedicated \region thus preventing control flow and stack tampering that could otherwise originate from the malicious ISR. While isolated, these ISRs remain interruptable allowing for nested interrupts.

If either of the aforementioned rules are violated, \acro triggers an exception, preventing \region execution from continuing. Since our prototype MCU, the MSP430, treats all exceptions with a device reset, we use the same mechanism. However, other types of (software-defined) exception handling are also possible. While resetting the device can impact availability, any \adv can already use run-time attacks to force device resets (e.g., by jumping to an invalid address among other exceptions). Thus \acro's exception handling does not provide \adv with more capabilities than already available.

\subsection{Adversary (\adv) Model}
\label{sec:adv_mod}

We assume an \adv that attempts to fully compromise the MCU software state. We assume that one or more \region resident programs contain vulnerabilities that enable control flow hijacks, ROP, and code injection attacks. Code external to \regions is assumed to be benign. We emphasize that being privileged does not imply being trusted. Thus, risky privileged code can be defined as untrusted in \acro.
\adv's goal is to exploit \region-resident and vulnerable code to compromise (otherwise benign) code outside \regions, by tampering with its control flow, program memory, or data. In other words, \adv aims to escalate a \region-resident vulnerability to compromise the rest of the system.  
Physical/hardware tampering attacks are out of the scope of this paper. In particular, we assume that \adv cannot modify/disable the physical hardware, induce hardware faults, or bypass \acro formally verified hardware-enforced rules.
Protection against physical \adv and hardware-invasive attacks is considered orthogonal and can be obtained via physical access control and standard tamper-resistance techniques~\cite{ravi2004tamper}. 

\subsection{\acrotitle Architecture}
\begin{figure}[h]
    \centering
    \includegraphics[scale=.35]{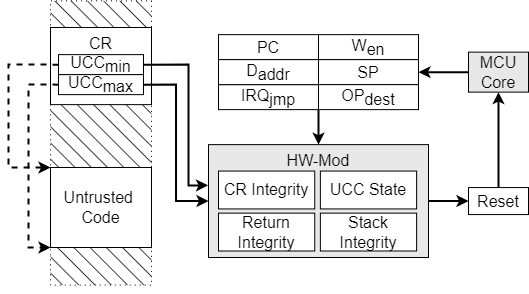}
    \caption{\acrotitle hardware architecture illustrating one \region}
    \label{fig:overview}
    \vspace{-1em}
\end{figure}

Figure \ref{fig:overview} depicts \acro's architecture. \acro adds a new hardware monitor, denoted HW-Mod, to the underlying MCU. \acro also reserves a dedicated region in memory to store \region configurations, i.e., \CR. \CR stores the address of each region's first and last instruction. The size of \CR varies with the number of simultaneous \regions supported. HW-Mod monitors the values within \CR to create isolated regions in memory. To detect violations, HW-Mod also monitors 6 additional signals from the MCU's core:
\begin{myitemize}
 \item The program counter (\PC), containing the address of the currently executing instruction.
 \item The data address access signal (\daddr), containing the memory address accessed by the current instruction (if any).
 \item The write enable bit (\wen), indicating if the current memory access (if any), is a write access.
 \item The stack pointer (\SP), indicating the memory address of the last data element added to the stack.
 \item The interrupt jump bit (\IRQ), indicating if a jump to an ISR is occurring.
 \item The operation return (\OP), containing the return address saved when call, interrupt, or exec instructions occur.
\end{myitemize}

If a violation of \acro properties occurs, a 1-bit \rst output signal is set. This signal resets the MCU core immediately, i.e., before executing the following instruction. As noted earlier, we treat violations with resets for simplicity but software-based exception handling is also possible. HW-Mod runs in parallel with the MCU core to monitor these values for each executed instruction. Table~\ref{tab:Notation} summarizes these signals and the notation used in the rest of this paper.

HW-Mod is composed of multiple sub-modules that enforce different \acro properties. The \CR Integrity sub-module protects \CR (which stores \region definitions) from being overwritten at run-time.
The Return Integrity sub-module enforces correct returns from \regions. The Stack Integrity sub-module prevents a \region from corrupting the stack pointer or overwriting external data in the MCU stack. Finally, the UCC State sub-module determines whether a \region is executing. This state is used by the Return and Stack Integrity sub-modules. A dedicated instance of the UCC State, Return Integrity, and Stack Integrity sub-modules is required for each isolated \region.

\begin{table}[]
    \centering
    \scriptsize
    \caption{\acro Notation}
    \vspace{-1em}
    \begin{tabular}{|c|p{0.77\linewidth}|}
         \hline
         Notation & Description\\
         \hline
         \region & Untrusted Code Compartment: An untrusted memory region\\
         \hline
         \regmin & The address of the first instruction of \region\\
         \hline
         \regmax & The address of the final instruction of \region\\
         \hline
         \CR & Configuration Region: Protected memory region that stores \regmin and \regmax for each \region\\
         \hline
         \PC & The current value of the Program Counter \\
         \hline
         \daddr & The memory address accessed by an MCU memory access \\
         \hline
         \wen & A 1-bit signal set when the MCU is writing to memory \\
         \hline
         \SP & The memory address of the current top of the stack \\
         \hline
         \IRQ & A 1-bit signal set if a jump to an interrupt is occurring \\
         \hline
         ISR & An Interrupt Service Handler executed for a given interrupt\\
         \hline
         \OP & The return address of call, interrupt, and exec instructions \\
         \hline
         \rst & A 1-bit signal indicating a violation occurred and resetting the MCU \\
         \hline
         $reset_{ucca}$ & A copy of the \rst signal used by each sub-module \\
         \hline
         \RTN & The expected return address of \region saved by \acro\\
         \hline
         \BP &  The address of the bottom of \region's isolated stack frame \\
         \hline
         
    \end{tabular}
    \vspace{-1em}
    \label{tab:Notation}
\end{table}

\subsection{\acro vs. Existing Hardware}
\label{mpucca}

As discussed in Section~\ref{sec:intro}, some MCUs have MPU support to protect memory regions. Therefore, a natural path to obtain untrusted code compartmentalization is with this existing support. Current MPUs enable the configuration of read, write, and execute permissions for up to 16 physical memory regions \cite{mpu_deetz}. These permissions are further split for privileged and unprivileged software, however, unprivileged code cannot have more permissions than privileged code~\cite{mpu_deetz, clements2018aces}. 

To isolate untrusted code, the MPU must first separate the untrusted code from the rest of the program. This can be done by setting the untrusted code as unprivileged and the remainder of the binary as privileged. Then the privileged code can be marked executable in privileged mode while the unprivileged (untrusted) code is executable in both contexts. This allows the application to freely call the untrusted code but prevents the untrusted code from jumping back into the rest of the binary. However, this does not prevent untrusted code from accessing other untrusted regions. As all untrusted code is unprivileged and executable by unprivileged code, independent untrusted segments can freely call each other, preventing isolation between untrusted regions. Similarly, the remainder of the application is now privileged. As privileged code can overwrite the MPU (and other system-level) configurations, this greatly increases the system's TCB. Also as the MPU only supports two privilege levels, isolating untrusted code prevents the MPU from isolating security-critical system code from applications in general.

While this model achieves isolation of untrusted code, it also prevents its execution outright. Since untrusted code is unprivileged it cannot jump back into the now privileged application, thus execution cannot return from the untrusted region. Remedying this requires an ``exit region" to handle these transitions. This privileged region needs to be executable to unprivileged (untrusted) code and all unprivileged return instructions must be instrumented to jump to the exit region. The exit region must also enforce return integrity. However, this requires saving the return address when calling untrusted code. As such, all branch instructions that could call untrusted code (including all dynamic branches) must also be instrumented. 

For stack integrity, the MPU must define another region around the current stack when entering an untrusted region and mark it as read-only to unprivileged code. Moreover, the MPU would also need to maintain a shadow stack \cite{burow2019sok, dang2015performance} of return addresses and protected stack definitions otherwise when untrusted regions call each other, the current return address and protected stack region would be overwritten, re-exposing the system to an attack.

Due to these requirements, implementing a single MPU \region would require at least 4 MPU regions. It also requires heavy binary instrumentation and dynamic MPU reconfiguration leading to increased run-time overheads. Isolating multiple regions further requires the implementation of a shadow stack. MPU-based \regions would also require disabling interrupts when executing \region-resident code. Otherwise, \adv could leverage interrupts to break isolation as they are privileged~\cite{inter_pri}. Thus MPU-based untrusted code isolation results in large run-time and storage overheads as well as precludes applications' real-time response to asynchronous events.

One could also attempt to port TrustZone controls into an untrusted code isolation mechanism. However, similar to the MPU case, this would also have many limitations. A TrustZone-based implementation would require all untrusted code to be in the Non-Secure world, while the rest of the application would execute in the Secure world. This would greatly increase the Secure world TCB. Similarly, this configuration would prevent TrustZone from isolating security-critical code from the rest of the application. Again similar to MPU, TrustZone cannot mutually isolate different untrusted code sections alone. Instead, Trustzone-equipped MCUs often work alongside an MPU to provide further separation within each world. However, this requires the MPU be reconfigured between worlds, increasing the system's run-time overhead. 
Similarly, all calls to and returns from (including interrupts) untrusted code will require execution to change worlds. This requires a context switch where the Secure world's state is saved/restored and the MPU configuration is updated before execution continues. Along with this, any Secure world data passed to untrusted code must be marshaled (copied) to the Non-Secure world and any results must be marshaled back. All this saving, copying, and configuring greatly increases the run-time overhead of the system.

\section{\acrotitle Details: Formal Specification and Verified Implementation}
\label{sec:methodology}

We now discuss \acro in detail. Our discussion focuses on a single \region as multiple \regions are obtained by simply instantiating multiple units of the same hardware modules (one per additional \region). To formally verify \acro's implementation, we formalize each of \acro's security properties using LTL. We then design Finite State Machines (FSMs) to enforce these requirements. The individual FSMs are implemented in Verilog HDL and combined into one Verilog design for HW-Mod (as shown in Figure~\ref{fig:overview}). Finally, HW-Mod and each sub-module are automatically translated to the SMV model checking language~\cite{mcmillan1993smv}, using Verilog2SMV~\cite{irfan2016verilog2smv}. The resulting SMV models are checked against all required LTL specifications, using the NuSMV model checker~\cite{cimatti2002nusmv}, to produce a proof of the correctness of \acro's implementation with respect to the LTL specifications.

\acro modules are implemented as Mealy FSMs (where outputs change with the current state and current inputs). Each FSM has one output: a local $reset$. \acro's output $reset$ is given by the disjunction (logic {\it or}) 
of the local $reset$-s of all sub-modules. Thus, a violation detected by any sub-module causes \acro to trigger an 
immediate MCU reset. To ease presentation, we do not explicitly represent the value of 
the $reset$ output in our FSMs. Instead, we define the following implicit representation:
\begin{myenumerate}
\item $\rst$ is 1 whenever an FSM transitions to the $Reset$ state;
\item $\rst$ remains 1 until transitioning out of the $Reset$ state;
\item $\rst$ is 0 in all other states.
\end{myenumerate}
Note that all FSMs remain in the $Reset$ state until $PC=0$, which signals that the MCU reset routine is finished.

\subsection{Defining Isolated \regions}
\label{iso}

Each \region is defined by the first and last addresses of its code: \regmin and \regmax, respectively. They mark the untrusted executable's location in memory. 
While \region can have arbitrary size, the smallest unit of code \acro can isolate is a single function, where \regmin and \regmax are the addresses of the first and last instruction in the function respectively. Attempts to isolate smaller regions (i.e., partial functions) would result in return integrity violations. Also, \regions should not partially overlap, since each \region is an independent code section. As such, partially overlapping regions would again cause return integrity violations. While partially overlapping \regions are invalid, \acro allows nested \regions. Nested \regions support different levels of distrust within an untrusted compartment, further constraining vulnerabilities within the inner \region from spreading to the outer region.
Similarly, each \region must be self-contained, i.e., include the untrusted code and its dependencies (such as callback implementations it relies upon). All other/trusted code should remain outside \region limiting its exposure to the potentially vulnerable code within \region. 

\subsection{Integrity of \region Boundaries}\label{sec:CR_integrity}

\regmin and \regmax can vary depending on the untrusted executable being compartmentalized. 
During cross-compilation/linking, appropriate \region values are determined and stored in \CR at load time. At run-time, \acro uses the values stored in \CR to monitor the execution of \region-resident code. 
To prevent \adv from altering \regmin and \regmax at run-time (effectively disabling \acro), \acro's \CR integrity sub-module ensures \CR is immutable. \CR integrity is defined in LTL specification~\ref{eq:ltl_CR} which states that at all times ({\bf G} LTL quantifier) \acro sets $\rst=1$ if an attempt to write to \CR is detected. Attempts to write to \CR are captured by checking if the \wen bit is set while the \daddr signal points to a location within \CR reserved memory. This ensures that \region definitions cannot be changed at run-time.
\begin{equation}\label{eq:ltl_CR}
\scriptsize
{\bf G:}\{[(D_{addr} \in CR) \wedge W_{en}] \implies reset\}
\end{equation}
\vspace{-1.5em}

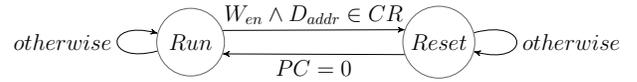
\begin{figure}[!hbtp]
\begin{center}
\scalebox{.6}{
    \begin{tikzpicture}[->,>=stealth',auto,node distance=5.5cm,semithick,
                    state/.style={circle, draw=black!60, minimum size=1.5cm},
                    node/.style={font=\large}]
                    \tikzstyle{every node}=[font=\Large]
    
    \node[state] (A){$Reset$};
    \node[state] (B)[left of=A]{$Run$};

    \path[->,every loop/.style={looseness=8}] (A) edge [loop right] node {$otherwise$} (A)
    (B) edge [loop left] node {$otherwise$} (B);
    
    \draw (A.200) -> node [below] {$PC=0$} (B.340);
    \draw (B.20) -> node [above] {$W_{en} \wedge D_{addr} \in CR$} (A.160); 
\end{tikzpicture} 
}
\caption{Verified FSM for \CR Integrity}
\vspace{-1.5em}
\label{fig:cr_int}
\end{center}
\end{figure}

Figure \ref{fig:cr_int} depicts the Verilog FSM implemented by the \CR integrity sub-module and formally verified to adhere to LTL specification~\ref{eq:ltl_CR}. The FSM has two states: $Run$ and $Reset$. The $Run$ state represents the MCU's normal operation. If an attempt to write to \CR is detected the state transitions to $Reset$. 
The FSM remains in this state until the reset process has been completed (indicated by having $PC=0$) at which point 
the FSM transitions back to the $Run$ state.

\subsection{Enforcing \region Return Integrity}
\label{sec:retcon}

Return integrity prevents control flow attacks within \region from escalating to the rest of the system by ensuring that \region returns to the correct address (disallowing any jumps from within \region to an invalid external location). Since \region has to isolate at least one function, execution must enter \region through a call or interrupt (irq) instruction and leave through a return instruction. \acro leverages this behavior to provide return integrity, by saving the correct return address internally (\RTN) when \region is called. Then when execution returns from \region, \acro checks that the actual return address matches \RTN. \textcolor{black}{Figure \ref{LTL_ret} depicts the LTL specifications defined to enforce return integrity.}

\textcolor{black}{\acro saves the return address rather than protecting its value on the stack 
as return instructions assume that the return address is at the top of the stack when called. In benign circumstances, this holds as data on the stack is freed (``popped'') before a return. However as \region is assumed to be vulnerable, execution can jump directly to a return instruction bypassing the required ``pops''. Thus protecting the return address alone would not prevent this type of attack.}

\begin{figure*}[!hbtp]
\scriptsize
\fbox{
  \begin{minipage}{\textwidth}
    
    \begin{equation}\label{eq:rst_ret_save}
        {\bf G:}\{reset \implies [(\lnot (PC \in UCC) \wedge ({\bf X}(PC) \in UCC) \implies ({\bf X}(RET_{exp}) = OP_{ret}) \vee reset) {\bf W}(PC \in UCC)]\} 
    \end{equation}

    \vspace{-1.6em}
    
    \begin{equation}\label{eq:ext_ret_save}
      {\bf G:}\{\lnot (PC \in UCC) \wedge ({\bf Y}(PC) \in UCC) \wedge \lnot{\bf Y}(IRQ_{jmp}) \implies [({\bf X}(PC) \in UCC \implies ({\bf X}(RET_{exp}) = OP_{ret}) \vee reset) {\bf W}(PC \in UCC)]\} 
    \end{equation}

    \vspace{-1.6em}
    
    \begin{equation}\label{eq:ret_immut}
      {\bf G:}\{(PC \in UCC) \wedge \lnot({\bf Y}(PC) \in UCC) \implies [(({\bf X}(RET_{exp}) = RET_{exp}) \vee reset) {\bf W}(\lnot (PC \in UCC))]\}
    \end{equation}

    \vspace{-1.6em}
  
    \begin{equation}\label{eq:irq_ret_immut}
      {\bf G:}\{\lnot (PC \in UCC) \wedge ({\bf Y}(PC) \in UCC) \wedge {\bf Y}(IRQ_{jmp}) \wedge \lnot {\bf Y}(reset) \implies [({\bf X}(RET_{exp}) = RET_{exp}){\bf W}((PC \in UCC) \vee reset)]\}
    \end{equation}

    \vspace{-1.6em}
    
    \begin{equation}\label{eq:ret_int}
      {\bf G:}\{\lnot reset \wedge (PC \in UCC) \wedge \lnot({\bf X}(PC) \in UCC) \wedge \lnot IRQ_{jmp} \implies ({\bf X}(PC) = RET_{exp}) \vee {\bf X}(reset)\}
    \end{equation}
  \end{minipage}
}
\caption{Return Integrity Module LTL Specifications}
\vspace{-2em}
\label{LTL_ret}
\end{figure*}

To check that \region returns to the correct location, \acro must first save the correct return address. LTL~\ref{eq:rst_ret_save} and LTL~\ref{eq:ext_ret_save} specify how \RTN is saved. Both statements stipulate that when execution enters \region, \acro sets \RTN to the correct return address (\OP) otherwise the device is in an invalid state (\rst). Whether the execution is entering \region is determined by the current and next \PC values. The next value of \PC is represented using the LTL ne\textbf{X}t operator X(\PC). If the current value of \PC is outside \region and X(\PC) is within \region, execution is entering \region. \OP is the correct return address as \OP is the return address written to the stack by the MCU core. Both statements are also conditioned on W($PC \in UCC$). This states that this \RTN saving behavior is true \underline{until} execution enters \region (or always true should execution never enter \region). In other words, this behavior is only true for the next execution of \region. While both specifications are similar, LTL~\ref{eq:ext_ret_save} states that when \region finishes executing, the correct return address is saved the next time execution enters \region. Whether the execution of \region is finished is determined by the current and previous values of \PC and the previous value of \IRQ. Previous values are represented using the LTL \textbf{Y}esterday operator (i.e., Y(\PC)). The \IRQ signal indicates if a jump to an ISR is occurring. 
If \PC was previously within \region and is now outside \region, execution has left \region. If execution left \region (and this was not due to an interrupt: $\lnot Y(IRQ_{jmp})$), then \region has finished executing. Since this statement conditions the next execution of \region on the previous iteration, it guarantees that the correct return address is saved every time \region is called, except for its first execution. Instead, LTL~\ref{eq:rst_ret_save} ensures the proper return address is saved for this initial execution. LTL~\ref{eq:rst_ret_save} states that after a device reset, the next time execution enters \region, \OP is saved to \RTN. \acro always initializes in a reset condition. As such, at boot, this statement also applies. Taken together, LTL statements ~\ref{eq:rst_ret_save} and ~\ref{eq:ext_ret_save} ensure that \RTN stores the correct value whenever \region is called. 

\textcolor{black}{Once saved, \RTN must remain fixed until \region finishes executing to ensure that return integrity only allows valid return addresses. Thus \RTN is immutable while executing \region.} This property is defined in LTL~\ref{eq:ret_immut}. Entrance to \region is again determined using the current and previous value of \PC. If \PC is currently within \region and the previous value was outside \region, then execution has just entered \region. \RTN's immutability is captured by checking that the current value of \RTN always matches the next (X(\RTN)) while within \region (W($\lnot (PC \in UCC)$)). However, \region is interruptable so to ensure that \RTN remains correct, \RTN must also be immutable across interrupts. LTL~\ref{eq:irq_ret_immut} describes this behavior and states that, when execution leaves \region due to an interrupt and the device is not resetting ($\lnot Y(reset)$), \RTN is immutable until execution of \region resumes, or until a device reset occurs. Added together these two specifications ensure that once execution of \region begins, \RTN cannot change until it finishes or the device resets.

Finally, return integrity is described in LTL~\ref{eq:ret_int}. This specification states that when execution exits \region (not due to an interrupt), an exception (reset) is triggered unless the actual and saved return addresses match. Unlike specifications~\ref{eq:ext_ret_save} and~\ref{eq:irq_ret_immut}, exiting a region is detected using the current and next value of \PC. Specifically, execution is exiting the region if \PC is currently in \region and the X(\PC) is outside \region. Due to this, X(\PC) is the actual value of the return address. Therefore, \acro compares \RTN to X(\PC) and sets $reset$ = 1 if a violation is detected.

\begin{figure}[!hbtp]
\begin{center}
\scalebox{.8}{
\noindent\resizebox{\columnwidth}{!}{
\begin{tikzpicture}[->,>=stealth',auto,node distance=6cm,semithick, state/.style={circle, draw=black!60, minimum size=1.5cm},node/.style={font=\large}]
tikzstyle{every node}=[font=\Large]
    
    \node[state] (A){$Reset$};
    \node[state] (B)[below of=A]{$In$};
    \node[state] (C)[left of=B]{$Out$};
    \node[state] (D)[right of=B]{$IRQ$};

     \path[->,every loop/.style={looseness=8}]
     (A) edge [loop above] node {$otherwise$} (A)
     (A.west) edge [bend right=25] node [above left] {\shortstack{$PC = 0 \wedge$ \\ $\lnot(PC \in UCC)$}} (C.north)
     (C) edge [out=145, in=115, looseness=8] node [above] {$otherwise$} (C)
     (D.north) edge [bend right=25] node [above right] {$reset_{ucca} = 1$} (A.east)
     (D) edge [out=25, in=55, looseness=8] node [above] {$otherwise$} (D)
     (B) edge [out=295, in=325, looseness=8] node [below] {$otherwise$} (B)
     (C.70) edge [bend left=25] node [right] {$reset_{ucca} = 1$} (A.200);
     
     \draw (C.0) -> node [above] {$PC \in UCC$} (B.180);
     \draw (B.200) -> node [below] {\shortstack{$\lnot (PC \in UCC) \wedge$ \\ $\lnot IRQ_{jmp} \wedge PC = Ret_{exp}$}} (C.340);
     \draw (B.0) -> node [above] {$IRQ_{jmp}$} (D.180);
     \draw (D.200) -> node [below] {$PC \in UCC$} (B.340);
     \draw (B.80) -> node [right] {\shortstack{$(reset_{ucca} = 1)$ \\ $\vee (\lnot (PC \in UCC)$ \\ $\wedge \lnot IRQ_{jmp} \wedge$ \\ $\lnot (PC = Ret_{exp}))$}} (A.280);
     \draw (A.260) -> node [below left] {\shortstack{$PC = 0 \wedge$ \\ $PC \in UCC$}} (B.100);
     \draw (C.south) -| (-6,-8) -| (0,-8) node [below] {$PC \in UCC \wedge IRQ_{jmp}$} -| (6,-8) -> (D.south);
\end{tikzpicture} 
}
}
\caption{Verified FSM for Return Integrity}
\vspace{-2em}
\label{fig:cont_invo}
\end{center}
\end{figure}
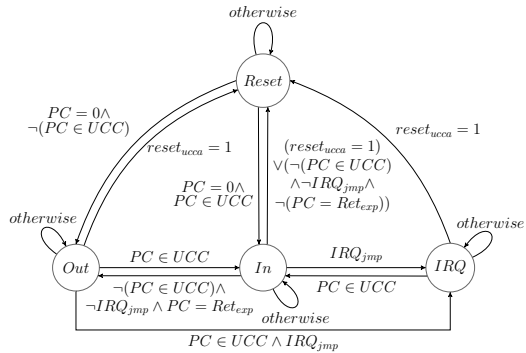

Figure \ref{fig:cont_invo} depicts the Verilog FSM implemented by the return integrity sub-module and formally verified to simultaneously adhere to LTL specifications \ref{eq:rst_ret_save}, \ref{eq:ext_ret_save}, \ref{eq:ret_immut}, \ref{eq:irq_ret_immut}, and \ref{eq:ret_int}. The FSM defines four states: $Out$, $In$, $IRQ$, and $Reset$. $Out$ represents when \PC is outside of \region. Once execution enters \region ($PC \in UCC$), the FSM transitions to $In$. While executing \region, the FSM remains in the $In$ state. If an interrupt occurs while within \region, the FSM transitions to the $IRQ$ state. If execution has just entered \region when an interrupt occurs, it is also possible for $Out$ to transition directly to the $IRQ$ state. $IRQ$ represents when \region has been interrupted. While in $IRQ$, \RTN is maintained. $IRQ$ transitions back to the $In$ state once \region resumes. When execution leaves \region ($\lnot (PC \in UCC)$) (not due to an interrupt), if execution returns to the expected memory address ($PC = RET_{exp}$) it is a valid return and the FSM transitions to the $Out$ state. Otherwise, a violation of return integrity has occurred and the FSM transitions to the $Reset$ state. Once the reset routine is completed, the FSM transitions to the $Out$ state. For synchronization, $Out$, $In$, and $IRQ$ also transition to $Reset$ if a violation occurs in another module or \region ($reset_{ucca} = 1$).

\subsection{\region Entry and Exit Points}
Despite being untrusted, \acro allows execution to enter and exit \region at/from any instruction in the region. This is because \acro prevents attacks within \region from escalating to the rest of the system. To that end, in terms of control flow integrity, it suffices to ensure that the \region caller code resumes correctly. As \region-resident code is untrusted (e.g., 3rd party libraries), \acro does not enforce properties regarding its internal behavior. By allowing arbitrary entry and exit points, multiple functions can be isolated by a single \region and all remain directly callable by external code.

\subsection{Protecting Stack Data Outside \region's Frame}
\label{sec:stackstackstack}

\begin{figure*}[!hbtp]
\scriptsize
\fbox{
  
  \begin{minipage}{\textwidth}
    
    \begin{equation}\label{eq:rst_bp_save}
      {\bf G:}\{reset \implies [\lnot({\bf Y}(PC) = PC) \implies (BP = {\bf Y}(SP)) \vee reset) {\bf W}(PC \in UCC)]\}
    \end{equation}

    \vspace{-1.6em}
    
    \begin{equation}\label{eq:bp_enter}
     {\bf G:}\{\lnot (PC \in UCC) \wedge ({\bf X}(PC) \in UCC) \implies ({\bf X}(BP) = BP) \vee reset\}
    \end{equation}

    \vspace{-1.6em}
   
    \begin{equation}\label{eq:ext_bp_save}
      {\bf G:}\{\lnot (PC \in UCC) \wedge ({\bf Y}(PC) \in UCC) \wedge \lnot{\bf Y}(IRQ_{jmp}) \implies [(\lnot({\bf Y}(PC) = PC) \implies (BP = {\bf Y}(SP)) \vee reset) {\bf W}(PC \in UCC)]\}
    \end{equation}
    
    \vspace{-1.6em}
    
    \begin{equation}\label{eq:bp_immut}
      {\bf G:}\{(PC \in UCC) \wedge \lnot({\bf Y}(PC) \in UCC) \implies [(({\bf X}(BP) = BP) \vee reset) {\bf W}(\lnot (PC \in UCC))]\}
    \end{equation}

    \vspace{-1.6em}

    \begin{equation}\label{eq:irq_bp_immut}
      {\bf G:}\{\lnot (PC \in UCC) \wedge ({\bf Y}(PC) \in UCC) \wedge {\bf Y}(IRQ_{jmp}) \wedge \lnot {\bf Y}(reset) \implies [({\bf X}(BP) = BP){\bf W}((PC \in UCC) \vee reset)]\}
    \end{equation}

    \vspace{-1.6em}
    
    \begin{equation}\label{eq:stack_iso}
      {\bf G:}\{[(PC \in \region) \wedge W_{en} \wedge (D_{addr} \geq BP)] \implies reset\}
    \end{equation}

    \vspace{-1.6em}
    
    \begin{equation}\label{eq:stack_integrity}
      {\bf G:}\{\lnot reset \wedge (PC \in UCC) \wedge \lnot({\bf X}(PC) \in UCC) \wedge \lnot IRQ_{jmp} \implies ({\bf X}(SP) = BP) \vee {\bf X}(reset)\}  
    \end{equation}
    \end{minipage}
    }
    
\caption{Stack Integrity Module LTL Specifications}
\vspace{-2em}
\label{LTL_stack}
\end{figure*}

Return integrity prevents escalation of attacks such as control flow hijacking and ROP. However, \region -resident code may still attempt to escalate data-flow attacks \cite{hu2016data,chen2005non, bellec2022rt} that overwrite data on the stack or create a malicious stack. Editing the stack has no immediate effect on a program's control flow. Therefore, return integrity is not violated. However, as a program's behavior depends on its variables, editing stack data could still compromise execution integrity.

To prevent data-flow attacks, \acro creates an isolated stack frame for \region. Stack frames are a memory management technique that segments the stack into different sections corresponding to different function calls \cite{one1996smashing}. To define a frame, \acro stores the initial stack pointer (\SP) of the previous instruction when entering \region. Since execution enters \region through either a call or interrupt, \acro saves the SP before the return address is pushed to the stack. We use this value for the base of the \region's frame for multiple reasons. First, this value separates non-\region and \region data. As no \region-resident code has been executed yet, all \region data will be written above this value. Second, this value is what \SP should be when execution returns from \region. When exiting \region, the return instruction removes the return address from the stack. Thus, upon exit, \SP should be the same value as before the call to \region. We refer to the saved \SP value as the base pointer (\BP) in the remainder of the paper. To isolate \region's frame, \acro blocks all write attempts performed by \region-resident code to addresses below \BP and enforces the proper stack context ($SP = BP$) when exiting \region.

Consequently, writes to stack variables passed by reference into \region are also blocked as they result in writes below \BP. Instead, as the heap is above the stack \cite{girard_2017} (and thus \BP) data passed by reference to \region should first be copied to the heap. Then when execution returns from \region, the edited heap value can be copied back to the original. We emphasize that (contrary to writes) the stack is always readable from \region. Thus this marshaling is only necessary for pre-existing stack data that is meant to be written by code within a \region. Global variables are also stored above the stack by default in the target architecture \cite{msp430usersguide, girard_2017} and thus writable by \region-resident code. This is expected as global variables are meant to be accessible to the whole program. Nonetheless, if desired, selected global data can be linked (at compile-time) to appear below the stack, preventing writes from \region.


Figure \ref{LTL_stack} lists the LTL statements defined to enforce stack integrity. LTL~\ref{eq:rst_bp_save} states that after a reset, whenever the executing instruction changes ($\lnot(Y(PC) = PC)$) \BP contains the previous value of \SP ($BP = Y(SP)$) until execution enters \region. While \BP saves the previous SP, this actually represents the initial \SP for the current instruction. Thus when \region is called, \BP holds the value of \SP before the return address is pushed to the stack. By conditioning on a reset, this statement ensures that \BP is correct when calling \region for the first time. LTL~\ref{eq:ext_bp_save} similarly states that when \region finishes executing, \BP stores the previous value of \SP whenever the current instruction changes until execution re-enters \region. This rule ensures the BP is also correct on all subsequent executions of \region. Once \region is running, LTL~\ref{eq:bp_immut} and~\ref{eq:irq_bp_immut} ensure \BP cannot be changed. LTL~\ref{eq:bp_immut} states that when execution enters \region, \BP is immutable until execution leaves \region. LTL~\ref{eq:irq_bp_immut} states that if \region is interrupted, \BP is immutable until \region resumes or a reset occurs. Together these statements ensure that once in \region, \BP cannot be changed until the execution of \region completes. \textcolor{black}{However, the value of \BP is ambiguous when execution enters \region. At this instance, LTLs~\ref{eq:rst_bp_save} and \ref{eq:ext_bp_save} do not hold, but, LTL~\ref{eq:bp_immut} only holds from this point forward. Thus to ensure \BP is still correct LTL~\ref{eq:bp_enter} states that when execution is entering \region, \BP does not change ($X(BP) = BP$). Combined with LTLs~\ref{eq:rst_bp_save} and \ref{eq:ext_bp_save}, these statements ensure that \BP is properly set whenever execution enters \region.}

\acro's stack frame isolation is defined in LTL specification \ref{eq:stack_iso}. This specification states that, at all times, \acro sets $\rst = 1$ if execution is within \region and attempts to write to the stack outside its stack frame. Writes outside the isolated frame are captured by the \wen bit being set while the \daddr signal points to a location below \BP. \daddr is below \BP if \daddr $\geq$ \BP as the stack grows towards $0$. Hence, values below \BP have a larger address than \BP. Stack isolation ensures that \region-resident code cannot tamper with data memory in use by the rest of the system. 

Finally, LTL~\ref{eq:stack_integrity} ensures that the stack pointer is properly restored before execution leaves \region. It states that, if the device is not already resetting and execution is leaving \region (not due to an interrupt), the next \SP should be \BP ($X(SP) = BP$). Since \BP represents the value of \SP at the start of the call to \region, this check enforces that \SP returns to the same value as before executing \region. This prevents an adversary from corrupting \SP such that malicious data written to the stack by \region resident code is used by non-\region code.

\begin{figure}[!hbtp]
\begin{center}
\scalebox{.8}{
\noindent\resizebox{\columnwidth}{!}{
\begin{tikzpicture}[->,>=stealth',auto,node distance=6cm,semithick,
                    state/.style={circle, draw=black!60, minimum size=1.5cm},
                    node/.style={font=\large}]
                    \tikzstyle{every node}=[font=\Large]
    
    \node[state] (A){$Reset$};
    \node[state] (B)[below of=A]{$In$};
    \node[state] (C)[left of=B]{$Out$};
    \node[state] (D)[right of=B]{$IRQ$};

     \path[->,every loop/.style={looseness=8}]
     (A) edge [loop above] node {$otherwise$} (A)
     (A.west) edge [bend right=25] node [above left] {\shortstack{$PC = 0 \wedge$ \\ $\lnot W_{en} \wedge$ \\ $\lnot(PC \in UCC)$}} (C.north)
     (C) edge [out=145, in=115, looseness=8] node [above] {$otherwise$} (C)
     (D.north) edge [bend right] node [above right] {\shortstack{$(reset_{ucca} = 1) \vee$ \\ $(PC \in UCC$ \\ $\wedge W_{en} \wedge$ \\ $D_{addr} \geq BP)$}} (A.east)
     (D) edge [out=25, in=55, looseness=8] node [above] {$otherwise$} (D)
     (B) edge [out=295, in=325, looseness=8] node [below] {$otherwise$} (B)
     (C.70) edge [bend left=25] node [right] {$reset_{ucca} = 1$} (A.200);
     
     \draw (C.0) -> node [above] {$PC \in UCC$} (B.180);
     \draw (B.200) -> node [below] {$\lnot (PC \in UCC)$} (C.340);
     \draw (B.0) -> node [above] {$IRQ_{jmp}$} (D.180);
     \draw (D.200) -> node [below] {$PC \in UCC$} (B.340);
     \draw (B.80) -> node [right] {\shortstack{$(reset_{ucca} = 1)$ \\ $\vee (PC \in UCC$ \\ $\wedge W_{en} \wedge$ \\ $\lnot D_{addr} \geq BP)\vee$ \\ $(\lnot PC \in UCC \wedge$ \\ $\lnot (SP = BP))$}} (A.280);
     \draw (A.260) -> node [below left] {\shortstack{$PC = 0 \wedge$ \\ $\lnot W_{en} \wedge$ \\ $PC \in UCC$}} (B.100);
     \draw (C.south) -| (-6,-8) -| (0,-8) node [below] {$PC \in UCC \wedge IRQ_{jmp}$} -| (6,-8) -> (D.south);
     
\end{tikzpicture} 
}
}

\caption{Verified FSM for Stack Integrity}
\vspace{-2em}
\label{fig:stack_pro}
\end{center}
\end{figure}
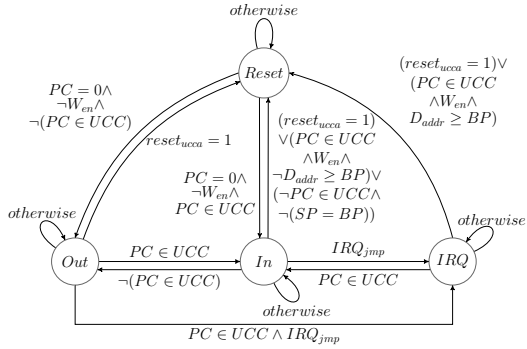

\textcolor{black}{Figure \ref{fig:stack_pro} depicts the Verilog FSM implemented by the stack protection module and formally verified to adhere to LTL specifications \ref{eq:rst_bp_save}, \ref{eq:bp_enter}, \ref{eq:ext_bp_save}, \ref{eq:bp_immut}, \ref{eq:irq_bp_immut}, \ref{eq:stack_iso}, and \ref{eq:stack_integrity}. The FSM defines four states, $Out$, $In$, $IRQ$, and $Reset$. The stack integrity FSM behaves similarly to the return integrity FSM with a few exceptions. Firstly, when in the $IRQ$ state, \BP is maintained until execution of \region is resumed rather than \RTN. Similarly, when transitioning to $Out$, \SP must equal to \BP otherwise the FSM transitions to the $Reset$ state. Finally, while in \region, any write below \BP will violate the stack isolation and cause the FSM to transition to the $Reset$ state.}

\section{Security Analysis}
\label{sec:eval}

Recall from Section \ref{sec:adv_mod} that \adv aims to escalate vulnerabilities located within \regions to compromise the rest of the system with attacks such as control flow hijacks, ROP, data corruption, and code injection. In this section, we argue that such attempts are unsuccessful due to \acro guarantees.

\adv may try to leverage vulnerabilities to alter the control flow of the binary and jump to an arbitrary location in memory. To do this \adv would need to exploit a branching instruction, such as a return, within a \region. To exploit a branching instruction, \adv would either need to overwrite a return/jump address on the stack or cause data on the stack to be misinterpreted as an address. Using vulnerabilities within \region, \adv could attempt to hijack an intermediate instruction or the final return instruction to jump to an arbitrary address.
However, this malicious jump would not match the saved return address (LTLs \ref{eq:rst_ret_save}, 
\ref{eq:ext_ret_save}, \ref{eq:ret_immut}, and \ref{eq:irq_ret_immut}) and the attack would be stopped (LTL \ref{eq:ret_int}).

\adv could also attempt to overwrite code in program memory or data on the stack. Both scenarios would allow \adv to alter the program behavior outside \regions. Program memory is located below the stack, thus always outside \region's isolated stack frame. Similarly, all data not belonging to \region falls below its frame's \BP, and is outside \region's isolated frame. As such, \adv cannot overwrite code in program memory and non-\region data on the stack (LTLs \ref{eq:rst_bp_save}, 
\ref{eq:bp_enter}, \ref{eq:ext_bp_save}, \ref{eq:bp_immut}, \ref{eq:irq_bp_immut}, and \ref{eq:stack_iso}). \adv could also attempt to write malicious data to the stack and then corrupt \SP such that the device uses the malicious stack once execution leaves \region. However, this would require \SP not be equal to \BP when leaving \region which is prevented by stack integrity (LTLs \ref{eq:rst_bp_save},
\ref{eq:bp_enter}, \ref{eq:ext_bp_save}, 
\ref{eq:bp_immut}, \ref{eq:irq_bp_immut}, and \ref{eq:stack_integrity}). Finally, \adv could attempt to inject and execute code on the stack or heap, however \acro prevents this as executing data memory requires execution to leave \region which violates return integrity (LTLs  \ref{eq:rst_ret_save}, \ref{eq:ext_ret_save}, \ref{eq:ret_immut}, \ref{eq:irq_ret_immut}, and \ref{eq:ret_int}).

Interrupts can bypass the isolation enforced by \acro. As such, \adv may try to abuse this behavior and exploit an interrupt to escape \acro's restriction. However, similar to any untrusted code in \acro, if an ISR is untrusted, it can also be defined as a \region. As a consequence, since the untrusted interrupt is isolated, return and stack integrity prevent it from escalating to the rest of the system (LTLs \ref{eq:rst_ret_save}, \ref{eq:ext_ret_save}, \ref{eq:ret_immut}, \ref{eq:irq_ret_immut}, \ref{eq:ret_int}, \ref{eq:rst_bp_save},
\ref{eq:bp_enter}, \ref{eq:ext_bp_save}, \ref{eq:bp_immut}, \ref{eq:irq_bp_immut},
\ref{eq:stack_iso}, and \ref{eq:stack_integrity}).
\adv could also attempt to overwrite the address of an ISR in the interrupt vector table (IVT). This would cause execution to jump to an \adv defined value, whenever the corrupted interrupt is triggered. Similar to program memory, IVT is stored below the stack. As such it is always outside \region's isolated stack frame and not writable by \adv (LTLs \ref{eq:rst_bp_save}, 
\ref{eq:bp_enter}, \ref{eq:ext_bp_save}, \ref{eq:bp_immut}, \ref{eq:irq_bp_immut}, and \ref{eq:stack_iso}).

Finally, \adv may attempt to disable \acro and break isolation by overwriting \region region definitions stored in \CR. However, \CR is immutable at run-time (LTL \ref{eq:ltl_CR}). The only way to overwrite \CR is by physically reprogramming the MCU which contradicts the \adv model.


\section{Prototype \& Evaluation}

We implemented \acro on the OpenMSP430 core~\cite{girard_2017}. \acro realizes the hardware architecture depicted in Figure \ref{fig:overview}. Along with HW-Mod, we implement a simple peripheral module for \CR. The peripheral module allows for \region definitions to be stored and accessed by HW-Mod at a pre-defined fixed data memory location. We use Xilinx Vivado~\cite{vivado} to synthesize an RTL prototype of \acro in real hardware. \acro's design was deployed on a Basys-3 prototyping board~\cite{digilent_ref}, that features an Artix-7 commodity FPGA~\cite{xilinx}. 
Our implementation is available at~\cite{anonymous642}.

\subsection{\acro Evaluation}

\noindent\textbf{TCB Size}. To calculate \acro's TCB size we count the amount of Verilog code needed to implement HW-Mod. Since \acro was implemented in hardware and works independently from the MCU core, \acro's TCB only consists of HW-Mod. The \acro prototype with support for 1 \region was implemented using \textbf{423 lines of Verilog code}. Each additional \region supported by \acro adds another 21 lines of Verilog to the TCB, for instantiating the same modules repeatedly.


\noindent \textbf{Hardware \& Memory Overhead}. The number of required \regions is application dependent. Due to this, we measure \acro considering support from 1 to 8 \regions and estimate the cost for arbitrarily many \regions. The additional hardware cost is calculated by looking at the number of added Look-Up Tables (LUTs) and Registers. The increase in the number of LUTs is 
an estimate of the additional chip cost and size required for combinatorial logic, while the number of registers offers an estimate of the state overhead required by the sequential logic in \acro FSMs. A summary of the hardware cost is shown in Figure \ref{tab:UCC_overhead}b. To isolate a single \region, \acro requires an additional 86 registers and 85 LUTs. This constitutes a respective $12.4\%$ and $4.7\%$ increase in registers and LUTs atop the unmodified OpenMSP430 core. In the largest test, with 8 \regions, \acro added 331 registers and 520 LUTs to the underlying system. This equates to a $47.8\%$ and $29\%$ increase in registers and LUTs.


In general, \acro can support arbitrarily many \regions with the only limiting factor being the additional hardware cost per region. We can predict the overhead for any \acro configuration as the overhead grows linearly with the number of \regions. As previously stated, \acro with one \region adds 86 registers to the MCU. However, each subsequent \region added only requires an additional 35 registers. Similarly, \acro with one \region adds an initial 85 LUTs to the MCU. Each additional \region adds on average 62 LUTs to the system (variance is due to the synthesis tool heuristic). Thus, \acro with support for $N$ \regions  can be estimated as:
\begin{equation}
\scriptsize
LUTs \approxeq 62\times(N-1) + 85
\end{equation}
\vspace{-1.75em}
\begin{equation}
\scriptsize
Registers = 35\times(N-1) + 86
\end{equation}

\acro also introduces a small storage overhead. Each \region's \regmin and \regmax are stored in \CR in the device's peripheral memory. Each address is 2 bytes long so each \region requires 4 bytes of data memory. On the OpenMSP430, peripheral memory can be between 512B and 32KB long~\cite{girard_2017}. Thus each \region incurs between $.01\%$ and $.78\%$ memory overhead depending on the size of peripheral memory.

\noindent \textbf{Energy Overhead}. To evaluate the energy consumption caused by \acro added hardware, similar to prior work \cite{noorman2013sancus, caulfield2023acfa, neto2023mathcal}, we use the Vivado synthesis tool~\cite{vivado} to estimate \acro's power consumption on our FPGA prototype. We consider \acro with support for 8 \regions. In this configuration, the MCU consumes 69 mW of static power with \acro accounting for 1 mW (1.45\%) of the total static consumption. The dynamic power consumption depends on how frequently \acro's internal registers are updated. We evaluate \acro on an application that loops through multiple function calls that modify the stack. We consider this a worst-case as it causes each \regions' internal \RTN and \BP to update constantly. Running this application resulted in a total dynamic draw of 113 mW where \acro accounted for 1 mW (0.88\%) of this consumption. Doubling the number of \regions to 16 increased the total dynamic draw to 114 mW. Thus, each \region introduces $\approx$ 0.125 mW of dynamic power draw.

\noindent \textbf{Run-time Overhead}. 
\acro does not modify the MCU core or Instruction Set Architecture (ISA). \region-related checks are performed by HW-Mod in parallel with the MCU core. These checks incur no extra run-time cycles to the software execution and thus do not interfere with the MCU's ability to respond to real-time events. As memory is accessed by HW-Mod through a dedicated physical channel, separate from the normal MCU core access channels, it does not cause interference or contention.

The only source of run-time overhead in \acro is due to marshaling data inputs to be modified by \region-resident code. In these cases, the data must be first copied to the designated heap region before calling \region-resident code. While the copying is done before \region execution, it affects the overall system run-time. 
The associated run-time depends on the amount of data to be copied.
In our prototype (based on MSP430), copying a ``word'' (2 Bytes, in this 16-bit architecture) requires one execution cycle of the absolute \texttt{MOV} instruction. This number scales linearly with the amount of data to be copied, i.e., an additional \texttt{MOV} instruction cycle is required for each pair of Bytes to be copied.

\noindent \textbf{Formal Verification}. We verified \acro on a Ubuntu 20.04 machine running at 3.70 GHz. Total verification time was about 11.5 minutes with maximum memory allocation of 125 MB, which is within the resources of commodity computers.

\noindent \textbf{Test Applications}. To demonstrate \acro protections, we implemented multiple test applications. These tests implement a simple user authentication program with a vulnerable input function, demonstrate several malicious cases that violate each of \acro's protections, and show how each attack is prevented. These tests are also available and discussed in more detail in our public \acro release~\cite{anonymous642}.


\subsection{Comparative Evaluation}

\begin{figure}
    \centering
    \begin{minipage}{.45\columnwidth}
        \centering
        \scalebox{.65}{%
            \begin{tikzpicture}
            \begin{axis}[
                xbar,
                xmin = 0,
                ytick = data,
                xlabel = {Additional Hardware Required},
                symbolic y coords ={CompartOS,Sancus, TrustLite,UCCA},
                nodes near coords,
                axis x line*=left,
                axis y line*=left,
                tickwidth = 0pt,
                height = 160pt,
                enlarge y limits = .2,
                xticklabels=none,
                scaled x ticks = false,
                every node near coord/.append style={font=\large},
                label style={font=\large},
                tick label style={font=\large},
                legend style={font=\large},
                x label style={at={(axis description cs:0.5,0.1)}},
                legend image code/.code={
                \draw [#1] (0cm,-0.1cm) rectangle (0.2cm,0.2cm); },
            ]
            \addplot [fill=orange] coordinates {
                (265,UCCA)
                (1145,TrustLite)
                (2366,Sancus)
                (10733,CompartOS)
            };
            \addplot [fill=darkgray] coordinates {
                (191,UCCA)
                (742,TrustLite)                    (1366,Sancus)
                (1588,CompartOS)
            };
            \legend{LUTs,Registers}
            \end{axis}
            \end{tikzpicture}
        }
        \vspace{-1.5em}
        \caption*{(a)}
    \end{minipage}
    \hfill
    \begin{minipage}{.4\columnwidth}
    \centering
         \scalebox{.68}{
            \begin{tabular}{|c|c|c|}
            \hline
            No. \regions & Registers & LUTs \\
            \hline
            1 & 86 & 85 \\
            2 & 121 & 145 \\
            3 & 156 & 205 \\
            4 & 191 & 265 \\
            5 & 226 & 327 \\
            6 & 261 & 389 \\
            7 & 296 & 450 \\
            8 & 331 & 520 \\
            \hline
            \end{tabular}
        }
        \vspace{1em}
        \caption*{(b)}
    \end{minipage}
    \caption{\acro Evaluation: (a) HW cost comparison with 4 \regions; (b) Added HW by total \regions}
    \label{tab:UCC_overhead}
    \vspace{-1.5em}
 \end{figure}
 
We compare \acro's overhead with three related schemes: Sancus~\cite{noorman2013sancus}, TrustLite~\cite{koeberl2014trustlite}, and CompartOS~\cite{almatary2022compartos}. 

Sancus provides memory isolation and attestation for shared remote embedded systems. Sancus introduces the \textbf{protect} and \textbf{unprotect} hardware instructions to create (and destroy) isolated software modules. Isolation is enforced by defining a fixed entry point for each module and by using the current program counter value to restrict access to a module's data to module resident code only. Sancus also enables key storage for each module to allow for remote attestation~\cite{eldefrawy2012smart,nunes2019vrased} of the region. 

TrustLite is another isolation architecture that isolates individual software tasks or trustlets. Trustlet definitions are recorded in the Trustlet Table in protected memory. For access control, TrustLite uses an Execution Aware MPU (EA-MPU) which extends the read, write, and execute permissions with the current value of the program counter. This allows the EA-MPU to restrict trustlet access to a predefined set of entry points and prevent access to trustlet data from outside the trustlet. The trustlets, Trustlet Table, and EA-MPU are all configured by a privileged process named the SecureLoader when the MCU boots.

CompartOS provides automatic software compartmentalization for high-end embedded systems. CompartOS uses the CHERI \cite{watson2020cheri} hardware capability system to create memory isolation. CHERI adds the capability data type and capability-aware instructions to the device's ISA. Capabilities extend integer pointers with metadata including bounds, permissions, and a validity bit to assign explicit permissions to the code they reference. Capabilities can also be "sealed" to link a code and data capability together and prevent their modification. CompartOS uses capabilities to define compartments and seals/unseals capabilities to context switch between different compartments.

We note that, while these approaches use hardware to isolate MCU memory, they are not directly comparable to \acro. None of the prior work focuses on isolating untrusted code sections, a feature unique to \acro. Both CompartOS and TrustLite target larger devices than \acro. \acro is more comparable to Sancus as both were implemented on the OpenMSP430 architecture. However, Sancus performs remote attestation in addition to isolation. Despite these differences, we believe that such systems are the most closely related to \acro. In our comparison, we consider default support for 4 isolated regions. The comparison is displayed in Figure \ref{tab:UCC_overhead}a.

\acro presents lower overhead. With support for four \regions, it requires 13.9\% of the registers and 12.1\% of the LUTs required by Sancus for the same number of isolated regions. With support for 8 \regions, \acro still only incurs about a fourth of the overhead (24.2\% registers and 22\% LUTs). \acro performs similarly when compared to TrustLite. \acro uses 25.7\% of the registers and 23.1\% of the LUTs TrustLite uses. At 8 \regions, \acro still only uses 44.6\% of registers and 45.4\% of LUTs used by TrustLite.

When compared to CompartOS, \acro uses 87.9\% fewer registers and 97.5\%  fewer LUTs to isolate four compartments. However, unlike Sancus and TrustLite, whose overhead scales with the number of isolated regions, CompartOS has the same hardware overhead, regardless of how many regions it supports. As \acro continues to isolate more regions, \acro's overhead will eventually surpass CompartOS's. However, these larger configurations are unlikely in low-end MCUs. Similarly, CompartOS uses 229\% more registers and 598\% more LUTs than the OpenMSP430 core itself. This overhead shows that CompartOS is impractical for such low-end MCUs.

\section{Extended Related Work}\label{sec:rw}
Aside from the techniques mentioned in Section~\ref{sec:intro}, there are several attempts to mitigate run-time vulnerabilities on MCUs.

\noindent\textbf{Control Flow Integrity (CFI)} is a class of techniques that limit the destination of any control flow transfer to a set of valid addresses~\cite{abadi2009control, serra2022pac, francillon2008code, davi2015hafix}. We also include randomization techniques in this discussion~\cite{shi2022harm, shacham2004effectiveness}. These approaches often use a Control Flow Graph (CFG) or a directed graph of nodes representing atomic sections of a binary~\cite{allen1970control}. CFGs enable the enumeration of all paths through a program, however, as programs get more complex the enumeration becomes undecidable. Due to this, many schemes use imprecise approximations prone to false positives~\cite{szekeres2013sok}. Other approaches focus solely on returns (notably, shadow stacks~\cite{dang2015performance}) removing the need for path enumeration but incurring large hardware and/or software overheads~\cite{szekeres2013sok}.

\noindent\textbf{MPU-based Compartmentalization} segments a binary into separate regions of memory and enforces isolation between them. Many schemes such as ACES~\cite{clements2018aces} simply use existing MPU operations to provide stronger isolation by segmenting code and enforcing well-defined entry points between them~\cite{clements2018aces,kwon2019uxom,kim2018securing,clements2017protecting,levy2017multiprogramming}. Other techniques extend MPU functionality by providing new isolation criteria~\cite{koeberl2014trustlite,brasser2015tytan,mera2022d}. For example, TrustLite~\cite{koeberl2014trustlite} uses the current instruction pointer to further restrict memory based on where the access originates or Toubkal~\cite{sensaoui2019toubkal} adds a new hardware monitor to restrict regions to specific hardware controllers such as cryptographic accelerators.

\noindent\textbf{ISA-based Compartmentalization} adds new functionality to the MCU core itself rather than relying on hardware monitors~\cite{almakhdhub2020mu,strackx2010efficient,noorman2013sancus,almatary2022compartos,xia2018cherirtos}. Self-Protecting Modules (SPMs)~\cite{strackx2010efficient} introduces two new hardware instruction to enable isolated memory regions. These controls also make use of the instruction pointer to validate memory accesses~\cite{noorman2013sancus}. Other controls also add new data types to the core such as CompartOS~\cite{almatary2022compartos} which adds the capability data type to the system along with the CHERI~\cite{watson2015cheri, watson2020cheri} ISA to use them. ISA-based isolation requires access to the source code to recompile the binary, with ISA-specific instructions. It also requires the CPU core and compiler to be trusted, increasing the system TCB and typically the hardware overhead. 

\section{Trade-Offs \& Limitations}\label{sec:lim}

\textbf{Fixed \region definitions and total number of \regions}. \acro implements \region definitions that are immutable at run-time. This enables \regions within privileged code and ensures \acro guarantees can not be disabled by any code at run-time. However, the total number of \regions can be limiting in larger systems with more untrusted code sections to isolate. In these systems, either untrusted code must share regions or not all untrusted code can be isolated. A trade-off would be allowing \region definitions to be configurable at run-time. This would allow for more flexibility and for \regions to be re-used by different code sections. However, it would introduce additional attack vectors and run-time overhead for switching the context between \regions. Alternatively, future work could further optimize the per-\region hardware cost in \acro, so that more \regions can be supported at the same cost.

\textbf{Protecting Heap data.}. By default, \acro does not prevent \region-resident code from accessing heap data. This design decision is based on the premise that many simple MCU applications avoid dynamic memory allocation for performance reasons. 
Nonetheless, in applications that require heap allocation, discretionary protection of heap data against \region-code can be achieved by linking a portion of the heap to allocate below the stack. This new portion would be protected from \region modifications (similar to how global variables are treated in \acro). 
A second unprotected portion of the heap could remain above the stack (where modifications can be made by \regions) and be used to share/marshal data into \regions. This approach allows selected heap data to be writable to \regions while protecting the rest of the heap and the stack. It also requires no changes to \acro hardware architecture.

\section{Conclusion}\label{sec:conclusion}
We proposed \acro: an architecture leveraging a formally verified hardware monitor to isolate untrusted code compartments (\regions) and limit the scale of run-time attacks on MCUs.
\regions are configurable and have variable size, making \acro compatible with different programs. Isolation of \regions is enforced in hardware and cannot be disabled by compromised software. In addition, \acro does not incur run-time overhead in terms of added CPU instructions/cycles. Similarly, \regions remain interruptable maintaining support for real-time operations. \acro's security analysis demonstrates that, by enforcing return and stack integrity for \regions, \acro constrains software exploits to their origin. Our evaluation, based on an open-source and formally verified \acro prototype, shows that \acro incurs small hardware overhead.

\bibliographystyle{IEEEtran}
\bibliography{references}

\end{document}